\documentclass[lettersize,journal]{IEEEtran}
\usepackage{graphicx}
\graphicspath{{Figures/}}
\usepackage{amsfonts}
\usepackage{amssymb}
\usepackage{amsmath}
\usepackage{color}
\usepackage{wasysym}
\usepackage{hyperref}
\usepackage{bbold}
\usepackage{algpseudocode}
\usepackage{placeins}
\usepackage{makecell}

\usepackage{mathtools}
\usepackage{multirow}
\usepackage{comment}
\usepackage{authblk}
\usepackage{datetime}

\definecolor{lightblue}{RGB}{136,163,209}

\begin{document}

\date{\today}

\title{Combinatorial optimization of connected UAV communication bridges for emergency response}

\author[1]{M. Vandelli}
\affil[1]{Quantum Computing Solutions, Leonardo S.p.A., Via R. Pieragostini 80, Genova, 16151, Italy}
\author[1,2]{D. Dragoni}
\affil[2]{Leonardo Hypercomputing Continuum, Leonardo S.p.A., Via R. Pieragostini 80, Genova, 16151, Italy}

\maketitle

\begin{abstract}
We present a combinatorial optimization problem for the strategic deployment of UAVs equipped with 5G antennas to assist rescue operations in regions hit by natural disasters. Our goal is to optimize the placement of UAVs to provide coverage in flying \emph{ad-hoc} networks among given candidate sites. Our formulation aims to maximize signal coverage and minimize interference while ensuring network connectivity. 
To mitigate interference effects, we incorporate the use of multiple frequencies. We formulate this problem as an integer quadratic program (IQP). We present numerical solutions obtained via the CPLEX solver and conduct a preliminary analysis of the problem's scalability in realistic network configurations. Our findings reveal a significant exponential increase in Time-to-Solution (TTS) as the number of sites grows, which poses a critical challenge in urgent, time-sensitive scenarios. To address this issue, approximate suboptimal solutions can be produced by enforcing a time limit on the solver. Although these solutions are not optimal, they preserve connectivity in most cases, providing a practical trade-off between solution quality and computational times that remain within feasible limits for real-time UAV redeployment. Recognizing the limitations of classical solvers in these contexts, we explore quantum computing as a promising alternative. Specifically, we reformulate the problem as a quadratic unconstrained binary optimization (QUBO) problem, suitable for most quantum algorithms. Through high-performance computing emulation, we show that the quantum adiabatic algorithm (QAA) can accurately solve small-scale instances, paving the way for future application of quantum computing to large-scale, time-critical optimization problems in disaster response.
\end{abstract}


\section{Introduction}

The aftermath of a natural disaster poses considerable challenges to the rescue-and-relief operations. Events such as earthquakes, hurricanes, and floods often lead to the severe damage or complete destruction of conventional communication infrastructure. This disruption hinders rescue operations, coordination efforts, and the dissemination of critical information, thereby exacerbating the crisis \cite{denning2006hastily} and ultimately impeding further relief actions such as the allocation of resources where they are most needed \cite{YI2007660, Jain2023}. Therefore, there is an urgent need for innovative solutions to quickly reestablish secure, low-latency, and high-capacity communication services in regions impacted by natural disasters. 

{Unmanned aerial vehicles (UAVs) have emerged as valuable assets for this purpose \cite{greenwood2020flying, https://doi.org/10.1002/rob.22075, 8641424, electronics12041051, CHANDRAN20241}, particularly due to their capability to operate in unknown or hazardous environments without endangering human personnel. Due to their independence from ground infrastructure, rapid deployment capabilities, and flexibility they can serve as relay stations (RS) to establish emergency communication links to rescuers on the ground, thereby providing \emph{on-demand} coverage to areas where traditional infrastructure has been partially or totally compromised \cite{BUSHNAQ2022159}.} 

In these situations, UAV swarms can be deployed to the affected region to create on-the-spot flying \emph{ad-hoc} communication networks (FANETs) \cite{BEKMEZCI20131254, 9045408, 9391631, 8167124} that do not rely on any pre-existing infrastructure, similar to wireless ad-hoc networks (WANETs) \cite{toh1997wireless}. This is accomplished by establishing a network backbone that provides connectivity between each UAV and the control-and-command center using multi-hop protocols. When pre-existing networks are available, FANETs can also enhance quality-of-service (QoS) by serving as a complementary network \cite{1207767}. These networks can leverage 5G millimeter-wave (mmWave) technology to offer low-latency and high-bandwidth connectivity, which are crucial for search-and-rescue operations \cite{8644135}.

Although \emph{ad-hoc} networks typically cover smaller areas compared to others based on satellite communications (SATCOM), they do not require additional communication hardware, resulting in lighter weights and longer battery life. Also, these networks are rapidly deployable, cost-effective for localized coverage, and suitable for real-time applications due to lower latency, making them ideal for emergency situations. While their range is limited, adding more nodes can extend coverage, though this increases complexity and potential interference \cite{10.1145/1080810.1080815}.

From a practical standpoint, the efficient deployment and operation of UAVs in disaster scenarios present substantial challenges. These include determining precise and safe UAV routes to targets \cite{thuy2022deploymentuavsoptimalmultihop, 10198846}, conducting careful mission planning \cite{act11010004, 7011909, 10620789}, ensuring low-latency data flow \cite{1673240}, positioning the UAVs in an optimal way \cite{ZHU2022103563, 6564778, Vandelli2024}, and dynamically reallocating the UAVs in response to real-time changes in the operational environment \cite{1036113}. 

In this work, we focus on the problem of optimal UAV placement, partially addressing the dynamic reallocation. Indeed, UAV deployment requires careful optimization to ensure sufficient coverage levels and guarantee a multi-hop path for data transfer to a central unit \cite{SUN2023, 7486987}, where peripheral information are gathered,  processed via high-performance computing (HPC) resources, and decisions are taken. 
\begin{figure*}[t]
    \centering
\includegraphics[width=\textwidth]{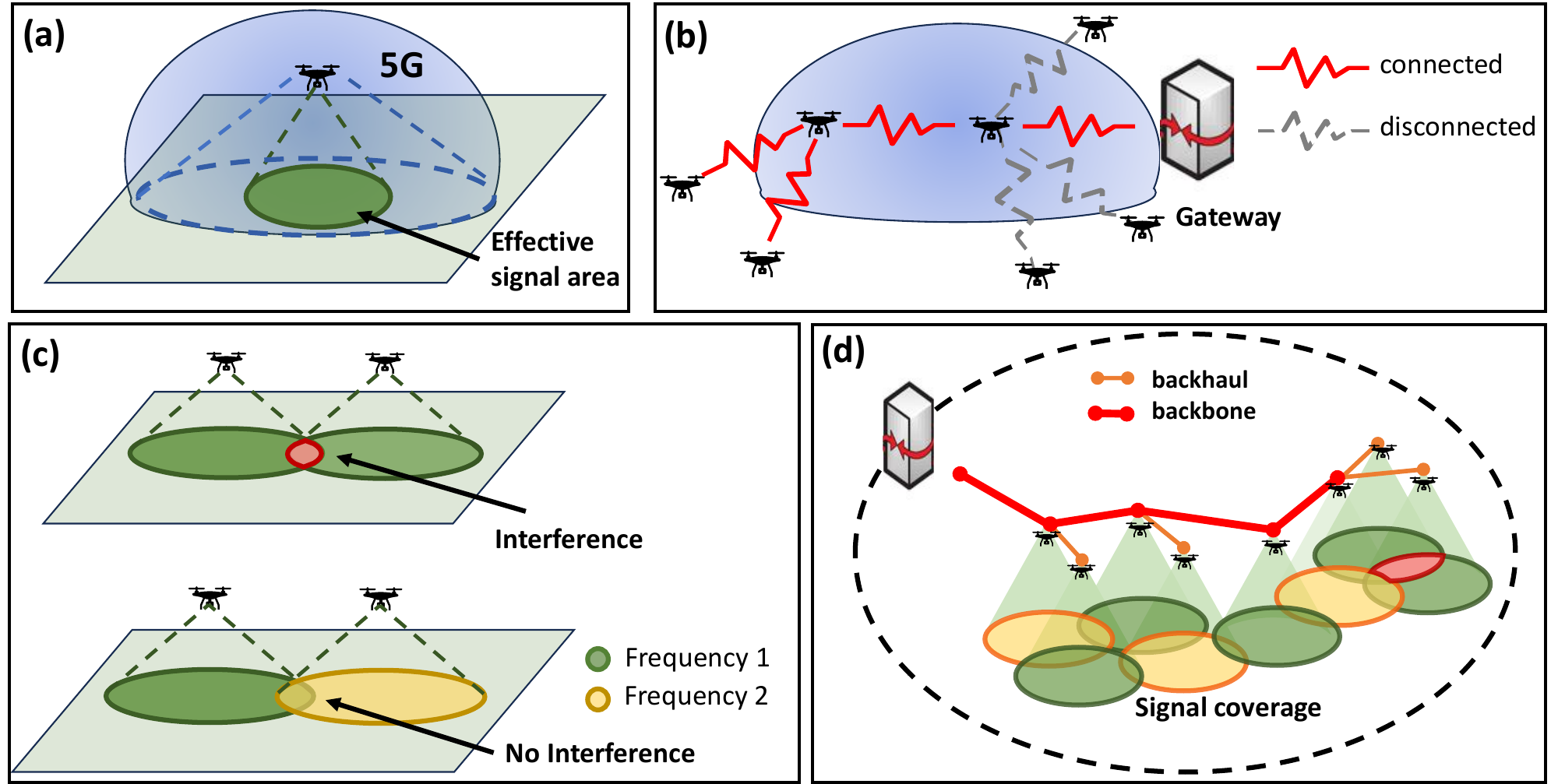}
    \caption{Description of the connected \emph{ad-hoc network} of UAVs modeled in this work. In panel (a), we show the basic building block of the network, a UAV equipped with a 5G antenna that provides coverage to the ground. The blue blob describes the full shape of the signal approximated as a sphere, while the green area shows the effective coverage provided by the UAV. Panel (b) shows the connectivity of the network. UAVs within the same blob can receive and transmit air-to-air signal to each other. Red wiggly pulses describe the backbone network, while orange ones indicate links that end there. Gray pulses indicate that it is not possible to establish a connection link between the two UAVs. Panel (c) shows interference between overlapping effective areas. Panel (d) shows the backbone assignment for a specific network with 2 frequencies. In particular, it highlights the distinction between backbone nodes connected by red lines and leaves connected to the backbone by orange lines.}
    \label{fig:model_scheme}
\end{figure*}
Our contribution to the field is the introduction of a novel combinatorial optimization problem, termed as \emph{Connected-Backbone UAV Placement} (c-BUP) problem. This problem focuses on deploying a swarm of UAVs to establish \emph{ad-hoc} networks, enabling communication between the rescuers moving on a territory. The resolution of this problem ensures connectivity of the UAV network and optimal signal coverage on the ground, while avoiding interference.

Specifically, we propose an integer quadratic program (IQP) to tackle this problem. Interference is encoded in a quadratic binary cost function, offering more flexibility than previously reported schemes based on circle packing \cite{mozaffari2016efficient}. We extend the model presented in Ref. \cite{Vandelli2024} to include both backbone nodes and the option to select multiple frequencies to reduce interference.

Additionally, our problem formulation can be directly mapped onto a QUBO problem, making it suitable for quantum computing solutions with  minimal overhead. Consequently, we explore solving this problem using both commercial classical solvers and the quantum adiabatic algorithm (QAA) \cite{farhi2000quantum, PhysRevA.67.022314, doi:10.1126/science.1057726} for small instances with a statevector emulator.

\section{Formulation of the Problem}

In this section, we describe the operative scenario and we introduce the IQP formulation of the problem. The main ingredients of our operative scenario are described in Fig.\ref{fig:model_scheme}. We select $N$ candidate sites on a territory. These sites are fixed and characterized by their geographic coordinates, and are obtained with a rough preliminary analysis of the territory: they could either be arranged on a regular grid or selected according to population, orography and/or other criteria. The number of available drones considered here is $V \leq N$, and each drone placed at site $v$ is equipped with two communication channels. The first is an \emph{air-to-ground} channel pointing towards the ground providing the signal to the rescuers, while the second is an \emph{air-to-air} channel used to transfer information among the drones themselves, and all the way to the entry point (or gateway), as depicted in Fig. \ref{fig:model_scheme}(a) and \ref{fig:model_scheme}(b) respectively. In this work we assume directional antennas for the air-to-ground channel and omnidirectional antennas for the air-to-air channel.
By deploying the UAVs at similar altitudes, we can enhance line-of-sight (LoS) connectivity among the UAVs and between the UAVs and ground personnel. This configuration enables the implementation of low-latency, high-throughput 5G transmission protocols, thereby improving communication efficiency and performance.

Our objective is to maximize the coverage area provided by the UAVs. However, coverage alone is not a sufficient indicator of optimal distribution of the UAVs. Indeed, ground coverage provided by neighboring drones might yield to regions of double coverage generating detrimental effects for the network performances such as signal interference and data transfer redundancy/reduced throughput \cite{10293450}, as indicated in panel (c) of Fig. \ref{fig:model_scheme}. To limit such effects we act in two ways. 

First, in our model we penalize the overlap of service areas generated by individual UAVs. Second, recognizing that some overlap in coverage areas is unavoidable due to the complex nature of the operational environment, we adopt a multi-frequency model. Distinct service frequencies are utilized in overlapping regions to prevent interference and maintain the service quality. 

This model is motivated by two assumptions: (1) the overlaps between signals at different frequencies do not cause significant interference, and (2) the personnel located on the ground is equipped with devices that transfer/receive data at a single frequency, thus avoiding data-exchange redundancy.

Besides maximizing coverage and reducing interference, we must ensure the connectivity of the network. Specifically, the signal from each UAV must travel through a connected high-speed, high-bandwidth path to reach the network's entry point, which is linked to the command-and-control center. To achieve this, we propose strategically deploying UAVs at specific sites to establish a network backbone. These UAVs ensure that each UAV is either part of the backbone or directly connected to at least one backbone UAV, as illustrated in panel (d) of Fig. \ref{fig:model_scheme}. UAVs not integrated into the backbone will be designated as \emph{leaves} in this connectivity graph, in analogy with tree graphs. In our model, we introduce a number $B \leq V$ of backbone nodes. The backbone UAVs may be specialized models with enhanced bandwidth and power capabilities, or standard drones positioned on certain sites to ensure connectivity. If the backbone consists of specialized UAVs, \( B \) denotes the number of such advanced drones available. Otherwise, \( B \) is a parameter that can be adjusted to optimize network performance. A smaller \( B \) reduces latency by minimizing the steps required to reach the gateway, but may limit coverage.  

\begin{figure}
    \centering
    \includegraphics[width=0.5\textwidth, trim={7.8cm, 1.2cm, 6cm, 1.4cm}, clip]{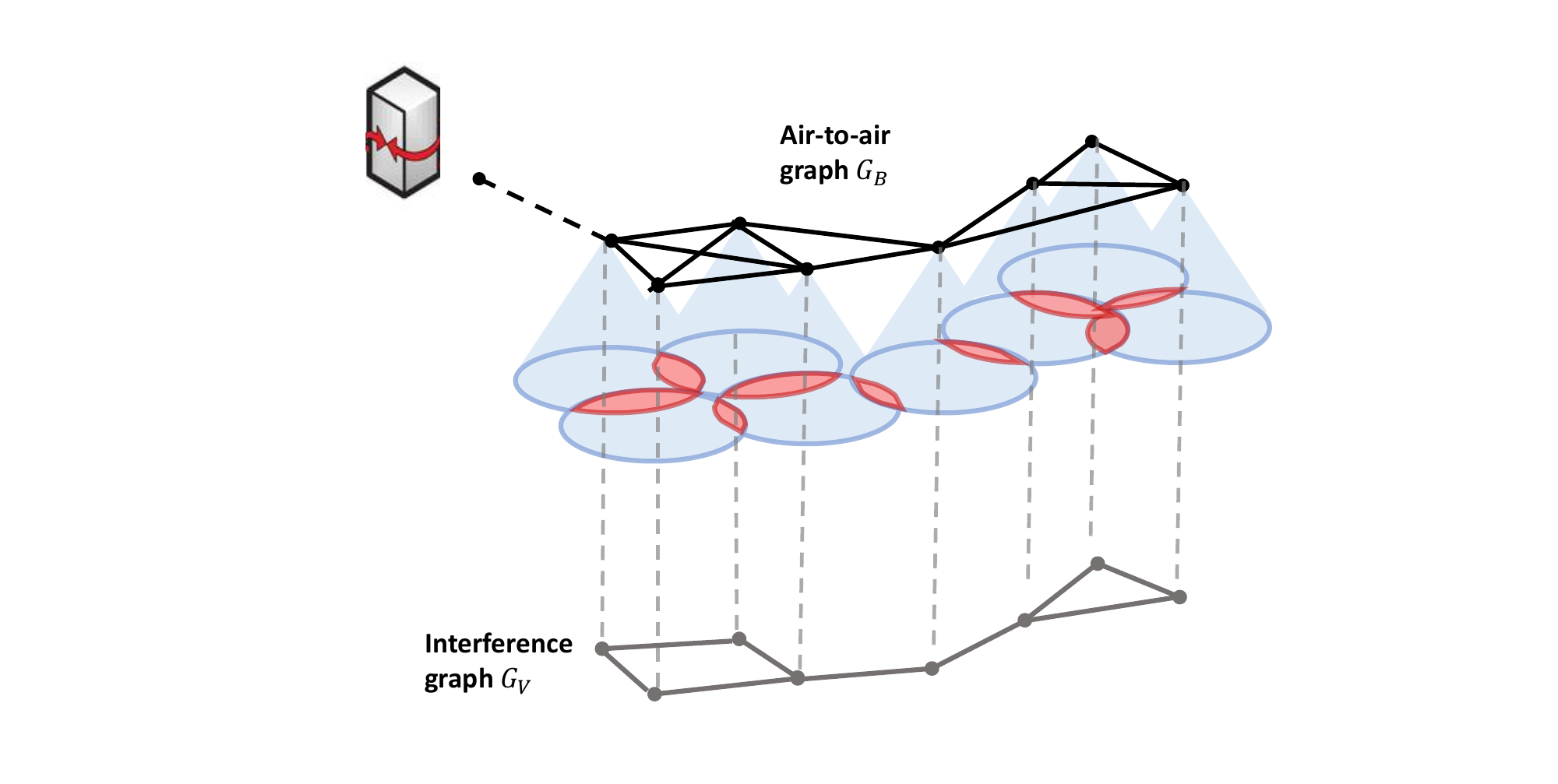}
    \caption{Illustration of the two graphs involved in the c-BUP problem. The air-to-air graph $G_B$ is shown at the top, while the interference graph $G_V$ is shown at the bottom of the picture. A solution of this problem with $V=N$ is shown as an example in Fig. \ref{fig:model_scheme} panel (d).}
    \label{fig:graphs_summary}
\end{figure}

In our model, the interference and the air-to-air UAV communication are represented by two graphs that share the same nodes corresponding to the potential sites, but can have different edges, as shown in Fig. \ref{fig:graphs_summary}. The two graphs are denoted as $G_B = (N, E_B)$, which represents the air-to-air connectivity between the UAVs, and $G_V = (N, E_V)$, which represents potential ground interference between the different antennas. Since the set of nodes is the same for both graphs, we define the set of nodes $\mathcal{N}=\{1,...,N\}$.
For the moment, we do not make assumptions on the relation between $E_B$ and $E_V$ in the mathematical formulation. Our model goes one step further with respect to previous works, which used the \emph{minimum connected dominating set} problem \cite{Sampathkumar, Levin2020} to model only the backbone of similar network problems \cite{10.1145/313239.313261, 5171125, 6180167}, by introducing a coupling between the optimization problems defined on the two graphs.

The resulting IQP is characterized by a quadratic cost function and several constraints. In the following, we break down the problem into into its main components: (\ref{sub:backbone}) creation of the backbone, (\ref{sub:multifreq}) multi-frequency coverage/interference optimization, (\ref{sub:coupling}) coupling between the two problems to ensure a consistent solution. In Sec.\ref{sub:summary}, we provide a recap of the full problem, while in Sec.\ref{sub:choice_B} we discuss the choice of the number of backbone drones $B$. Finally, in Sec.\ref{sub:qubo} we provide the QUBO formulation.

\subsection{Creation of the air-to-air backbone
\label{sub:backbone}}

We want to place $B$ backbone antennas on $N$ sites in such a way that they are connected. This amounts to finding a list of vertices $\{v_i\}$ with $i \in \{1, ..., B\}$ which are connected. The problem variables $x_{v,i}$ are binary variables with index the vertex $v \in \mathcal{N}$. If $x_{v,i}=1$ then $v$ is the $i-$th vertex in the sequence $\{v_i\}$. 
Unfortunately, there is no straightforward local way of fixing connectivity as a linear hard constraint. However, we can introduce this condition in the form of a cost function $H_B$ as
\begin{align}
  H_B(x) &= -{\sum_{(v,u) \in E_B}}\sum_{i=1}^{B-1} x_{v,i} \, x_{u,i+1} \, ,
  \label{eq:back_cost}
\end{align}
in which the binary string $x$ belongs to the space of bitstrings $\mathcal{S}_{N\times B}$. This term favors configurations in which each pair of consecutive vertices in $\{v_i\}_{i=1,...,B}$ are connected by an edge, effectively introducing connectivity of the backbone. 

For the sake of simplifying the notation, we introduce a variable $X_v$ that indicates whether site $v$ is assigned to the backbone as
\begin{align}
   X_v =  \sum_{i=1}^{B} x_{v,i}. 
   \label{eq:X_v}
\end{align}
The occupied nodes are those identified by $X_v = 1$. The set of backbone nodes is then defined as $\mathcal{B} = \{v \in \mathcal{N} \;| X_v = 1\}$. Clearly a solution is feasible only if $X_v \leq 1$.
We can also define the number of backbone antennas in the solution 
\begin{align}
  N_B(x) = \sum_{v=1}^N\sum_{i=1}^B x_{v,i} = \sum_{v=1}^N X_v
  \hspace{0.5cm} &\text{(backbone number)}.
\end{align}

With the definition \eqref{eq:back_cost}, the backbone solution can be found from the following IQP problem
\begin{align}
    x^* &= \arg\!\smashoperator{\min_{x \in \mathcal{S}_{N \times B}}} H_B(x) 
    \\
    &{\rm s.t.} \notag \\
    &X_v \leq 1  
\label{eq:back_ass_const}\\
    &\sum_{v=1}^N x_{v,i} = 1 {\hspace{0.5cm}} \forall i=1, ...,B\\
    & N_B(x)  = B. \label{eq:back_num_const}
\end{align}
The cost function gives negative weight if $(v,u)$ is connected by an edge in $E_B$ and no weight otherwise. 
The constraint \eqref{eq:back_ass_const}  makes sure that each node appears at most once in the sequence. This implies that $X_v$ is a binary variable. Strictly speaking this is not true anymore if we use Lagrange multipliers to convert hard constraints into soft ones, but we can tune the Lagrange multipliers so that it holds for the solution.
The second constraint is a number constraint on the number of backbone nodes.
Usually, there is a set $\mathcal{E}$ of possible entry points of the network, f.i. the boundary points of the territory. In this case, we introduce the condition
\begin{align}
    \sum_{v^* \in \mathcal{E}}X_{v^*} = \sum_{v^* \in \mathcal{E}}\sum_{i=1}^B x_{v^*, i} \geq 1.
    \label{eq:entry_points}
\end{align}
If the entry point $v^*$ of the backbone is known, then the set $\mathcal{E}$ contains a single point $ \{v^*\}$ and the condition simplifies to
\begin{align}
    X_{v^*} = \sum_{i=1}^B x_{v^*, i} = 1.
    \label{eq:entry_point}
\end{align}

\subsection{Multi-frequency antenna placement problem \label{sub:multifreq}}

The placement problem is similar to the one presented in Ref. \cite{Vandelli2024}. However, we include the possibility of having $F$ different operation frequencies that we can assign to each active site. 
We choose for simplicity a one-hot encoding for the frequency, although different schemes may lead to a lower number of variables. This is introduced as additional terms in the cost function, in a similar way as a (weighted) \emph{graph coloring problem} \cite{Lucas_2014}. 
To model frequency assignment, we introduce additional binary variables $y_{v,p}$ to each site where $v\in \mathcal{N}$ indicates the site index and $p \in \{0, 1, ..., F\}$ indicates if the site is empty (index $p=0$) or operates at frequency $\nu_p$. 

Additionally, we define the occupation of a site by an antenna as
\begin{align}
   Y_v = \sum_{\bf p > 0}^F  y_{u,p}, 
    \label{eq:Y_v}
\end{align}
while the summation with also $p=0$ included is denoted as 
\begin{align}
   Y^{\bf (0)}_v = \sum_{\bf p = 0}^F  y_{u,p}. 
    \label{eq:Y0_v}
\end{align}
The set of occupied sites is then identified as $\mathcal{V} = \{v \in \mathcal{N} \; |\; Y_v = 1\}$.

We consider the simple case with only overlap, coverage and a limited number $V$ of available UAVs.

The terms of this problem are
\begin{align}
    &N(y) = \sum_{v=1}^N Y_v \hspace{0.5cm} &\text{(number)}\notag\\
    &H_O(y) = \smashoperator{\sum_{\substack{p>0,\\p'>0 \\{(u,v) \in E_V}}}} O^{p,p'}_{uv} y_{up} y_{vp'}  &\text{(overlap)} \notag\\
    &H_C(y) = - \left(\sum_{v=1}^N A_v Y_v - \smashoperator{\sum_{{(u,v) \in E_V}}} A_{uv} Y_u Y_v\right) &\text{(coverage)} 
\end{align}
where $A_u = A(C^{\rm int}_u) = \pi (r^{\rm int}_u)^2$ and $A_{uv} = A(C^{\rm int}_u \cap C^{\rm int}_v)$, i.e. the area of the intersection of the two circles centered at sites $u$ and $v$.
It is important to consider only $p>0$ since $p=0$ means that the site is empty, so it doesn't contribute to the cost terms. The coverage term $H_C$ is approximated, since the quadratic term eliminated the double counting, but triple and higher-order overlaps are not eliminated. This approximation is reasonable in a low-density configuration of the sites, in which the connectivity is local. 
In this work, we consider for simplicity the case in which signal interferes only if the antennas operate at the same frequency, so $O^{p,p'}_{uv} = O_{uv} \delta_{p,p'}$ where $\delta_{p,p'}$ is the Kronecker symbol and $O_{uv} = \text{[Area of overlap] }$.

We then define the cost function of the site assignment as the sum of overlap and coverage contributions as
\begin{align}
    H_V(y) &= H_O(y) + H_C(y).
    \label{eq:reg_cost}
\end{align}

The binary program describing only the placement/frequency assignment becomes
\begin{align}
    y^* &= \arg\!\smashoperator{\min_{y \in \mathcal{S}_{N \times (F+1)}}} H_V(y) \\
    &{\text{s. t.}} \notag\\
    &N_V(y)  = V\\
    &Y^{(0)}_v = 1, \hspace{0.5cm} \forall v
    \label{eq:IQP_freqs}
\end{align}
where the constraints indicate that we have $V$ antennas and each site has only one possible operating state (color/frequency).

\subsection{Coupling between air-to-air graph and placement problem \label{sub:coupling}}

Now we introduce a coupling between the two problems. First, we define the set of the \emph{leaves} $\mathcal{L}$ as
\begin{align}
    \mathcal{L} = \mathcal{V} \setminus \mathcal{B} = \{v \in \mathcal{N} \;|\; Y_v = 1 \land X_v = 0 \}
\end{align}
as opposed to the backbone nodes $\mathcal{B}$.
The coupling terms between the two problems should enforce the following requirements:
\begin{enumerate}
    \item If a site is assigned to the backbone $v \in \mathcal{B}$, so $X_v = 1$, it must also have a non-zero frequency index $Y_v=1$. For this reason, we have to impose $\mathcal{B} \subseteq \mathcal{V}$.\label{en:1}
    \item The number of leaves $\mathcal{L}$ connected to backbone nodes is maximized. \label{en:2}
    \item Conversely, connections between a leaf node $v \in \mathcal{L}$ and a pair of backbone nodes $u, w \in \mathcal{B}$ are unfavorable, since it increases the redundancy in the exchange of information. \label{en:3}
\end{enumerate}

Point \ref{en:1} is imposed by introducing the onsite coupling term
\begin{align}
    H^{(0)}_{VB}(x, y) = -\sum_{v=1}^N X_v Y_v.
\end{align} 
Point \ref{en:2} is introduced in the form of a cost function. We want to maximize the number of antennas (active sites) that receive signal from the backbone.
\begin{align}
    H^{(1)}_{VB}(x,y) &= - \smashoperator{ \sum_{(v,u) \in E_V}} \left(Y_v X_u + Y_u X_v\right).
\end{align}
This formulation is the first term of the polynomial expansion that counts all the covered leaves. In order to make it more rigorous, we should remove double counting that stems from multiple connections.

\begin{figure}[t]
    \centering
    \includegraphics[width=0.4\textwidth]{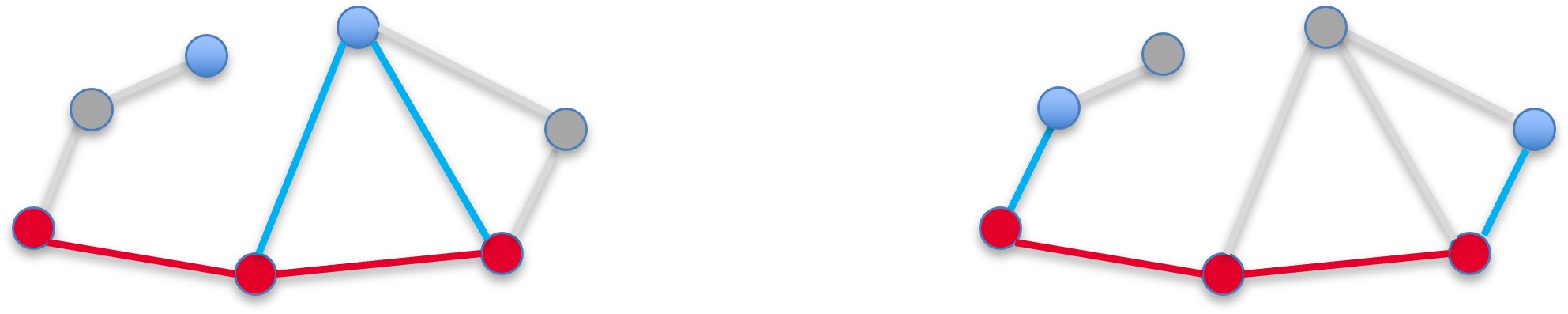}
    \caption{Example of the reason why we include the cubic $VB$ interaction term. The quadratic $VB$ term counts the number of connections only, so the two graphs above would have the same cost without $H^{(2)}_{VB}$, despite having lower connectivity between leaves and backbone nodes. The cubic term instead favors connection between a leaf node and \emph{a single} backbone node.}
    \label{fig:cubic_int}
\end{figure}

Indeed, point \ref{en:3} is obtained by computing the second term in the polynomial expansion. Effectively this term removes the double counting from double connections between two backbone nodes and one leaf. Explicitly, it reads
\begin{align}
    H^{(2)}_{VB}(x,y) &= \sum_{u=1}^N \sum_{(v,w) \in N_{V}(u)} Y_u X_v X_w
    \label{eq:H2_VB}
\end{align}
where $N_{V}(u)$ are the neighbors of vertex $u$ on the $G_V$ graph.
Since this term couples 3 variables at a time, we call it the cubic interaction.
This penalty term also removes the degeneracy between configurations as the one shown in Fig. \ref{fig:cubic_int}.

This term can be handled using solvers able to deal with cubic terms or it can be decoupled including slack a binary variable $z_{u; v, w}$ for each triple overlap. 
In principle, this decoupling requires $\mathcal{O}(N^3)$ slack variables for each triple of vertices $(u, v, w)$. However, since usually the connectivity is sparse and local, the number of slack variables is typically much smaller that that. A good way to compute the non-zero triples is to fix $y_u$ and compute the couples of nodes connected to $u$.
The IQP formulation with slack variables $z$ reads \cite{DINNEEN202360}
\begin{align}
    H^{(2)}_{VB}(x,y; z) &= \sum_{u=1}^N \sum_{(v,w) \in N_{V}(u)} \left[Y_u \, z_{u; v, w} + X_v \,X_w\right.\notag \\
    &\left.- 2 (X_v + X_w)\, z_{u; v, w} +3 z_{u; v, w} \right].
\end{align}
This decoupling works only because constraint \eqref{eq:Y_const} forces $X_v$ and $Y_v$ to be binary variables. 

We define the total coupling as
\begin{align}
    H_{VB}(x,y; z) = H^{(0)}_{VB}(x,y) + H^{(1)}_{VB}(x,y) +  H^{(2)}_{VB}(x,y; z).
\end{align}

\subsection{Full problem \label{sub:summary}}

To summarize, we define the total cost function of the problem as
\begin{align}
        H_{\rm full}(x, y; z) = &\lambda_V H_V(y) + \lambda_B H_B(x) + \lambda_{VB} H_{VB}(x,y; z).
\end{align}
We remind that, despite the definitions of Eq.\eqref{eq:X_v} and \eqref{eq:Y_v}-\eqref{eq:Y0_v} to simplify the notation, the binary variables defining the solution are still $x_{v,i}$, $y_{v,p}$ and $z_{u,v,w}$.

The overall problem becomes
\begin{align}
   (x^*,y^*&, z^*) = \arg\!\smashoperator{\min_{(x,y,z)}} H_{\rm full}(x,y; z)
   \label{eq:full_problem}\\
    &{\rm s.t.} \notag\\
    &Y^{(0)}_v = 1, \hspace{0.5cm} &\forall v \in \mathcal{N}\label{eq:Y_const}\\
    &X_v \leq 1 , \hspace{0.5cm} &\forall v \in \mathcal{N}\\
      &\sum_{v=1}^N x_{v,i} = 1, {\hspace{0.9cm}} &\forall i\in \{1, ..., B\}\\
    & N_B(x) = B,
    \label{eq:backbone_num}\\
    & N_V(y)  = V,\\
    &\sum_{v^* \in \mathcal{E}} X_{v^*} \geq 1 \label{eq:full_problem_const}
\end{align}
where the positive Lagrange multipliers satisfy $\lambda_B > \lambda_{VB} \gg \lambda_V$ to effectively force connectedness of all the components of the network. The number of variables in this formulation is $N_q = N (B + F + 1 ) + N_{\rm 3-over}$.

\subsection{Optimization of the number of backbone nodes \label{sub:choice_B}}

So far, we treated $B$ as a fixed parameter of the problem. However, if the backbone UAVs are not of a different model, $B$ effectively is free to vary. We can choose it based on the required performances in the specific use-case. 

In our scenario, we imagine that the number $B$ has to be minimized to reduce the latency of the network, since $B$ is effectively an upper bound on the number of data-exchange steps between the leaves and the gateway. We denote the best option as $B^*$. We envisage two ways of performing this optimization.

\begin{itemize}
    \item {\bf Direct optimization of $B$}:  In order to carry out the minimization exactly, we should minimize the problem \eqref{eq:full_problem} with cost function $H_{\rm full}(x,y; z) - \lambda_{NB} N_B(x)$ and replace the constraint \eqref{eq:backbone_num} with
\begin{align}
    N_B(x) \leq V.
\end{align}
This option requires to define $V$ backbone variables $x_{v, i}$ with $i=1, ..., V$. Since typically $V > B^*$, this option leads to a rather large number of variables. 

\item {\bf Iterative optimization of $B$}: Iteratively optimize $B$ based on some stopping criterion: either minimum value of $B$ that ensures connectivity of all the leaves or minimum cost with respect to $B$. This option allows to find the solution by solving multiple smaller IQPs compared to the direct option.
\end{itemize}

If $V=N$, then the problem is reduced to a connected dominating set problem with the additional condition of coverage maximization. If we additionally choose $B^*$ as the minimum $B$ that ensures connectivity, we recover the \emph{minimum connected dominating set} problem as a particular case, which is a known NP-hard problem \cite{FERNAU20116290}. On other hand, the placement problem with multiple frequencies is a combination of graph coloring problem and Ising problem on a non-planar graph in the general case. These two problems are known to be NP-hard \cite{Karp1972, FBarahona_1982}. As a result, we argue that also finding the solution of general c-BUP problem instances belongs to the NP-hard complexity class. We argue that decomposition techniques, such as the SPLIT framework \cite{vandelli2025parallelsplittingmethodlargescale}, could be used to reduce the computational burden of this problem, as done for a similar frequency allocation problem \cite{vandelli2025constraintpreservingquantumalgorithmmultifrequency}.

\subsection{QUBO formulation for quantum computation \label{sub:qubo}}

In order to solve this problem with quantum computers, we need to formulate it in the Ising form \cite{Lucas_2014, lodewijks2020mapping} or equivalently in the QUBO form \cite{Kochenberger2014}. The problem \eqref{eq:full_problem}-\eqref{eq:full_problem_const} does not have inequalities so the number of variables remains the same upon conversion to the unconstrained problem. Additionally, if we consider an algorithm that can handle Ising-like models with cubic terms, such as the QAA, we can reduce the number of variables to $N(B+F+1)$. In a gate-based model, the cubic terms can be decomposed into 2-qubit gates anyway \cite{PhysRevLett.91.027902}.

In that case, the QUBO model reads
\begin{align}
    Q(x, y; z) = \;&H_{\rm full}(x, y; z) \notag\\ &+\lambda_N \left[ (N_V(y) - V)^2 + (N_B(x) - B)^2 \right]\notag \\&+ \lambda_P \sum_{v=1}^N\left[\left(Y^{(0)}_v- 1\right)^2 + \left(X_v - z_v \right)^2\right] \notag\\&+ \lambda_P \sum_{i=1}^B\left[ \left(\sum_{v=1}^N x_{v,i}\right) - 1\right]^2
\end{align}
where we introduced the slack variables $z_v$ with $v \in \mathcal{N}$ to encode inequality constraints.
On a quantum annealer, we would still need to include the slack variables $z$ responsible for the decoupling of the cubic term, since the current devices only support quadratic models \cite{Willsch2022}.

\section{Model for realistic operative conditions}

\subsection{Realistic network parameters}

In emergency response scenarios, the use of Internet-of-Things support and predictive analytics necessitates a high data exchange rate, which can be ensured by 5G transmission technologies.
In this study, we specifically consider 5G transmission frequencies in the range 1 to 30 GHz, which typically achieve a LOS transmission range of a few hundred meters \cite{zhang2019survey}.
 
 As stated before, we consider UAVs all placed at roughly the same height equipped with omnidirectional 5G antennas for air-to-air connections \cite{8620298} and directional antennas for air-to-ground connections. In this case, each site $v$ is associated with two circles with maximum transmission and interference radii $r^{\rm int}$ and $r^{\rm back}$ respectively. We assume that a pair of nodes $v$ and $u$ has an air-to-air connection if their distance $d(u,v) \leq r^{\rm back}$ and they potentially interfere with each other if $d(u,v) < 2r^{\rm int}$. A significant interference occurs if the previous condition for a polynomial-in-$N$ pairs of sites, requiring the solution of the c-BUP. 
We assume here the UAVs operating at an altitude of approx 150 m and equipped with directional antennas pointing to ground generating an effective coverage cone of 45-degree angle, although actual operative conditions may vary depending on the specific type of application. 
Our choices result in an maximum interference radius of \( r^{\rm int} \approx 150 \) m, meaning that two UAVs will interfere with each other if they are within 300 meters of one another. To ensure backbone connectivity while considering interference events, we assume the connectivity radius to be in the range \( r^{\rm back} = 150 \sim 400 \) m, which is a realistic maximum distance to maintain a sufficient connectivity to ensure high data-transfer rates \cite{10198846}. This leads to a range $r^{\rm back}/r^{\rm int} = 1.33 \sim 2.66$. Values of $r^{\rm back}/r^{\rm int} < 2$ correspond to a $G_B$ more dense than the $G_V$, while $r^{\rm back}/r^{\rm int} > 2$ corresponds to the opposite case.

\begin{comment}
\begin{table}[h!]
\centering
\begin{tabular}{ |c|c|c|c||c||c| } 
\hline
$h$ & $\theta$  &  $i$  & $r^{\rm int}$  & $r^{\rm back}$ & $r^{\rm back}$/$r^{\rm int}$ \\ 
\hline
\hline
     &    &       &      &  50  & 0.33 \\ 
 150 & 45 & 212.1 & 150  &  200 & 1.33 \\ 
     &    &       &      &  500 & 3.33 \\  
\hline
     &    &       &      &  50  & 0.50    \\ 
 100 & 45 & 141.4 & 100  &  200 & 2.00    \\  
     &    &       &      &  500 & 5.00    \\  
\hline
     &    &       &      &  50  & 1.00 \\ 
 50  & 45 & 70.7  & 50   &  200 & 4.00 \\ 
     &    &       &      &  500 & 10.0 \\ 
\hline
\hline
     &    &       &      &  50  & 0.58 \\ 
 150 & 30 & 173.2 & 86.6 &  200 & 2.31 \\ 
     &    &       &      &  500 & 5.80  \\  
\hline
     &    &       &      &  50  & 0.87    \\ 
 100 & 30 & 115.5 & 57.7 &  200 & 3.47    \\  
     &    &       &      &  500 & 8.70    \\  
\hline
     &    &       &      &  50  & 1.73 \\ 
 50  & 30 & 57.7  & 28.9 &  200 & 6.90 \\ 
     &    &       &      &  500 & 17.3 \\ 
\hline
\hline
\end{tabular}
\caption{Distance in meters and angles in degrees. Note that if the transmission technology/protocol is the same for A2A and A2G antennas then we expect the condition $r^{\rm back}$/$r^{\rm int} \geq 1$ to hold (A2A is LOS, while A2G is not). Note also that the condition $r^{\rm back}$/$r^{\rm int} = 2$ discriminates between the two regimes where the interference and A2A connectivity are denser than the other.}
\label{table_scenarios}
\end{table}
\end{comment}

\subsection{Site location and complexity \label{sub:site_loc}}

In this subsection, we discuss the complexity of our model when applied to realistic UAV parameters. While the connectivity problem itself is not computationally challenging, its complexity arises from the coverage-interference problem $H_V(y)$. This problem becomes intricate when the underlying graph lacks symmetry between the sites.
The inhomogeneity between sites can arise from two primary sources. The first involves variations in the interference graph, where the sites are arranged on a non-regular grid, leading to spatial disparities in overlap events between different sites, i.e. the coefficients $A_{uv}$ and $O_{uv}$ do not possess symmetries. The second source stems from unequal weights $A_v$ assigned to each site, which occurs when coverage is influenced by non-uniform factors such as device distribution, probabilistic constraints, or varying priority levels across regions. Both of these scenarios can be mapped onto a spin-glass model with an inhomogeneous local magnetic field, a class of problems that, in the most general case, is known to be NP-hard \cite{FBarahona_1982}. Conversely, the case of regular grid with equal covered areas turns out to be trivial, since a solution can be easily found by placing the backbone nodes on a straight line and then saturating the connections.

In our calculation, we assume a scenario in which the candidate sites $\mathcal{N}$ are arranged on a non-regular grid. 
We imagine to construct these points based on the distribution of rescuers on the territory and exploiting a theorem from continuous optimization, which states that the candidate sites needed to cover all the devices are the Circle Intersection points \cite{CHURCH1974101, BLANCO2021105310}. If some points lie at a distance larger than $r^{\rm back}$ from all the other points, we can introduce additional points to appropriately connect them to the rest of the graph (f.i. by bisection). We assume that we have applied some pre-selection technique to reduce the number of sites to a reasonable amount.

\section{Numerical solution of selected instances}

\subsection{Details of the simulations}

We simulate the site pre-selection described in Subsec.\ref{sub:site_loc} by generating $N$ nodes on a $1 \; {\rm km} \; \times 1 \; {\rm km}$ area using pseudo-random Sobol' sequences with Lloyd's optimization scheme. This procedure ensures that the points are roughly evenly distributed on the square but do not possess any particular symmetry, hence they constitute an irregular grid, as it is the case for real instances. This also ensures that the instances cannot be solved using trivial schemes.
Here, we fix $r^{\rm back}/r^{\rm int} = 2.2$, which leads to significant overlaps in the generated graphs while ensuring sufficient backbone connectivity.

In these simulations, we treat $B$ as an hyperparameter of the problem. For this reason, we identify the generated instances with a string $(N, V, B, F)$. All the solutions of the IQP problem are obtained with the free version of the IBM CPLEX solver \cite{cplex2022v22}. To simplify and streamline the generation of the optimization problems, we used the \texttt{docplex} python software library, while the graphs are generated using the \texttt{networkx} library.

\subsection{Effect of the cubic term on solution connectivity}

\begin{figure}[ht!]
    \centering
\includegraphics[width=0.5\textwidth]{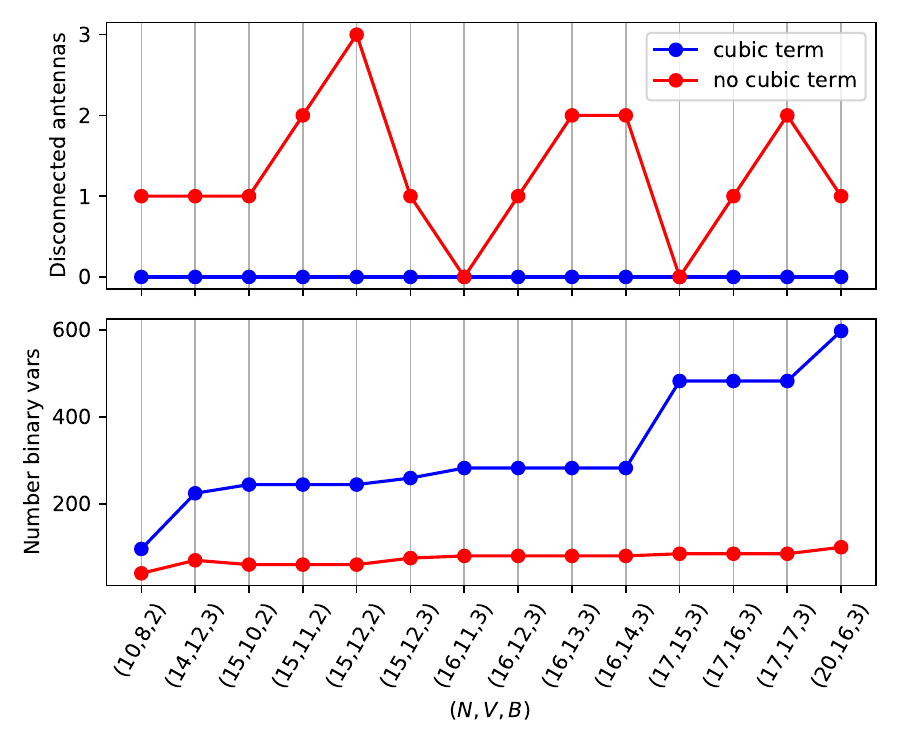}
    \caption{Number of leaves disconnected from the backbone in the presence or absence of the cubic term for several configurations. This plot shows the importance of that term for the overall connectivity. All the configurations have been chosen in such a way that a fully connected solution is feasible, meaning that the number of connections is smaller than $C_{\rm max}$.}
    \label{fig:cubic}
\end{figure}

We first investigate the effect of the $H^{(2)}_{VB}$ term of Eq.\eqref{eq:H2_VB}. In the upper panel of Fig. \ref{fig:cubic}, we show the connectivity of the solution in the presence (light blue dots) and absence (orange dots) of the cubic term of Eq.\eqref{eq:H2_VB} for several instances of the problem. We see that the presence of the cubic term effectively enforces an enhanced connectivity of the problem in all the cases considered here. The lower panel shows the the variable overhead of including the cubic term in a case with local connectivity. The cubic term requires additional slack variables $z$ to use integer quadratic program solvers such as CPLEX, leading to much larger time-to-solution (TTS) compared to the case without $H^{(2)}_{VB}$. We argue that, in cases that require a quick solution, the model without cubic terms could give sufficiently accurate solutions combined with heuristics to attach the disconnected leaves to the backbone as a second step.

\subsection{Examples of solutions}

\begin{figure}[ht!]
    \centering
\includegraphics[width=0.5\textwidth]{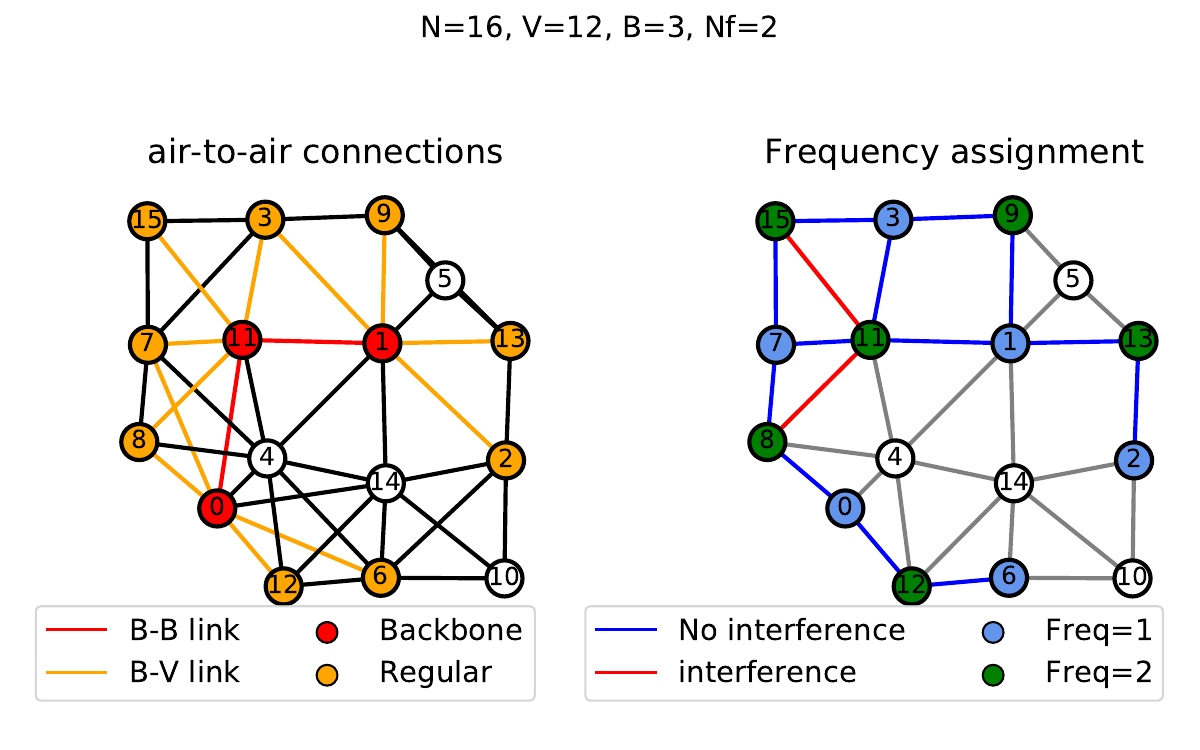}
    \caption{Solution of the problem for the instance $(16, 12, 3, 2)$. On the left, we represent the connectivity, while to the right we show the frequency assignment. In this case with $F=2$, we notice two interference conflicts.}
    \label{fig:example_nf2}
\end{figure}

\begin{figure}[ht!]
    \centering
\includegraphics[width=0.5\textwidth]{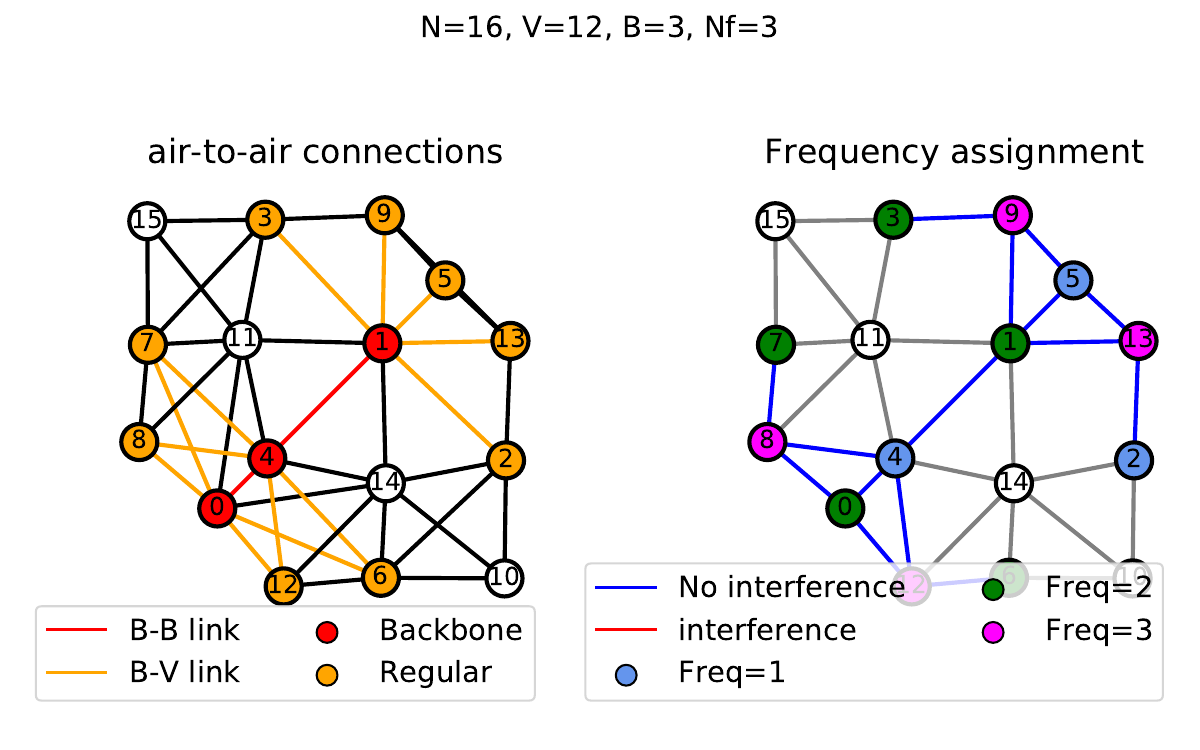}
    \caption{Solution of the problem for the instance $(16, 12, 3, 3)$. On the left, we represent the connectivity, while to the right we show the frequency assignment. In this case with $F=3$, we notice that all the interference is avoided.}
    \label{fig:example_nf3}
\end{figure}

\begin{figure}[ht!]
    \centering
\includegraphics[width=0.5\textwidth]{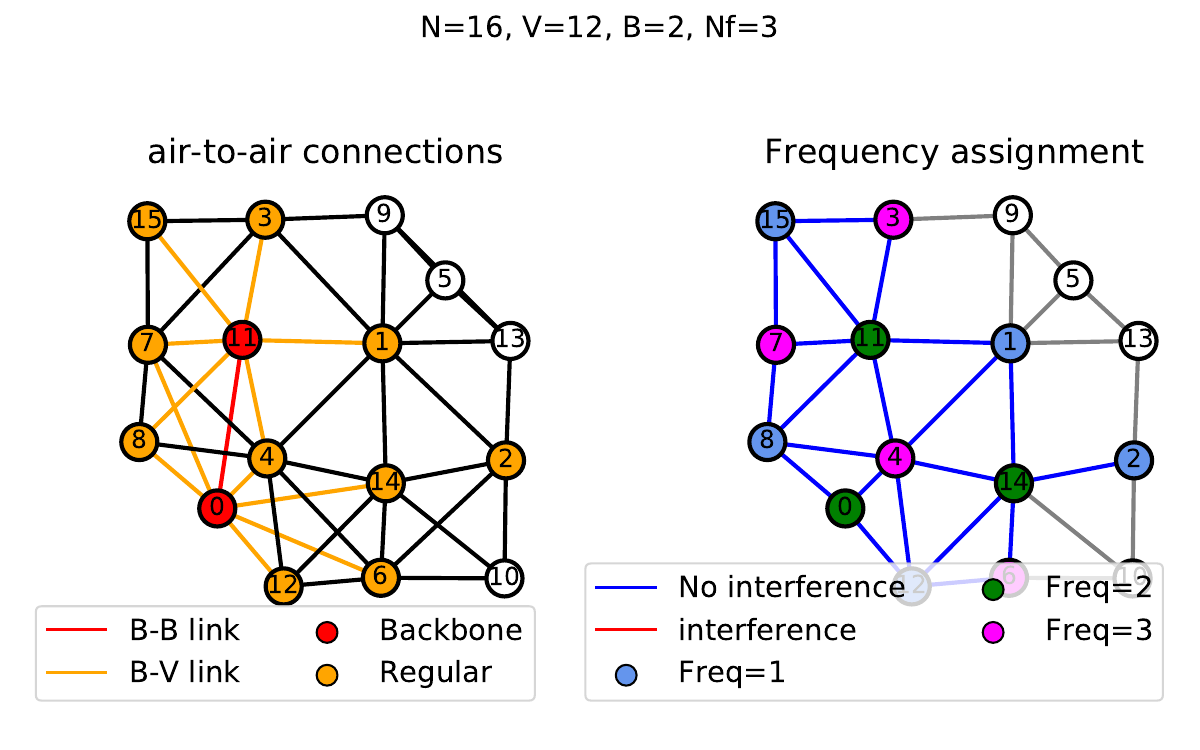}
    \caption{Solution of the problem for the instance $(16, 12, 2, 3)$. On the left, we represent the connectivity, while to the right we show the frequency assignment. In this case with $B=3$, we see that it is not possible to connect all the leaves to the backbone. Specifically, $v=5$ is assigned but disconnected.}
    \label{fig:example_nf3_bsmall}
\end{figure}

Here, we discuss the two examples of solutions shown in Fig. \ref{fig:example_nf2} and \ref{fig:example_nf3}. The two problems are characterized by the same number of nodes $N=16$, $V=12$ and $B=4$. However, they have different number of frequencies, $F=2$ and $F=3$ respectively.
In both figures, we show the air-to-air connection graph $G_B$ (left graph) and the interference graph $G_V$ (right). Since $r^{\rm back}/r^{\rm int} > 2$, the edges of the two graphs are different and specifically $E_V \subset E_B$. The solution is superimposed to the graphs. In particular, solution connectivity is displayed on the $G_B$ graph, while frequency assignment and interference are associated with the $G_V$ graph. For what concerns air-to-air connections, white circles indicate empty sites, while red and orange dots indicate the backbone nodes and leaves respectively. Red links indicate connections between two backbone nodes (B-B), while orange links show links between backbone nodes and leaves. Moving to the frequency assignment graph, blue links indicate a connection between antennas operating at different frequencies while red links indicate equal operating frequency. Only the latter results in interference. The different colors of the nodes show their assigned frequency. 

In both problem instances, we see that the number of backbone nodes is sufficient to ensure full connectivity of the leaves to the backbone. However, in Fig. \ref{fig:example_nf2}, we observe that $F=2$ frequencies are not sufficient to eliminate all the interference, while $F=3$ avoids all the interferences. The reason for this is that anytime a triangle is formed, the only way to avoid all interferences is to assign each vertex to a different frequency. In more densely-packed graphs, removing all the sources of interference in general involves an increasingly large number of frequencies. The fact that, by just changing the frequency, the backbone changes shows the feedback of frequency assignment on connectivity.

\begin{figure}
    \centering
    \includegraphics[width=0.5\textwidth]{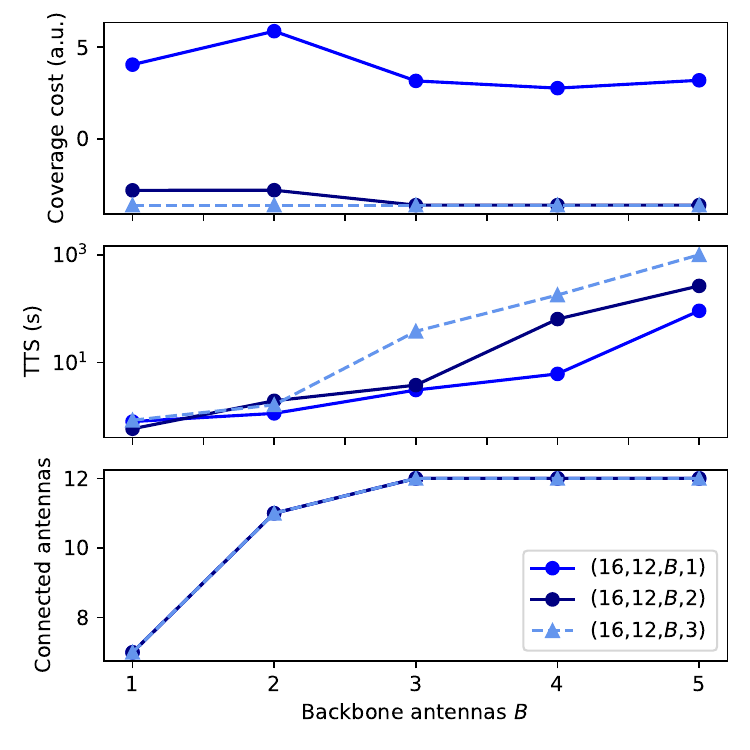}
    \caption{Figures of merit for the instances characterized by $N=16$ and $V=12$ \emph{vs} the backbone size $B$. Different colors and markers indicate different number of frequencies $F$. Top panel shows the cost of coverage/interference $H_V(y^*)$, the middle panel shows the TTS and the lower panel shows the number of connected leaves.}
    \label{fig:cost_vs_B}
\end{figure}

Finally, in Fig. \ref{fig:example_nf3_bsmall} we show that choosing $B=2$ is not enough to ensure connectivity of the solution since node $v=5$ is disconnected. Hence, the minimum $B^*$ that ensures full connectivity is 3. The crucial metrics are summarized in Fig. \ref{fig:cost_vs_B}. We see that the coverage cost $H_V(y)$ computed for the solution $y=y^*$ is much higher for $F=1$ signaling a large number of interference events. We know that the opposite case $F=3$ is sufficient to eliminate all the overlaps, as shown in Fig. \ref{fig:example_nf3}. In the intermediate case $F=2$, the selected overlaps are so small that for $B \leq 3$ the difference with respect to $F=3$ is smaller than the marker size. For what concerns the TTS, it comes as no surprise that it grows faster than exponentially with $B$. In the lower panel, we show the number of connected antennas, defined as the size of the largest connected subgraph of $\mathcal{V}$. This quantity increases monotonically with $B$ and converges to the total number of antennas $V$ at $B=3$.

\subsection{Numerical analysis of selected instances}

\begin{figure}[t]
    \centering
    \includegraphics[width=0.5\textwidth]{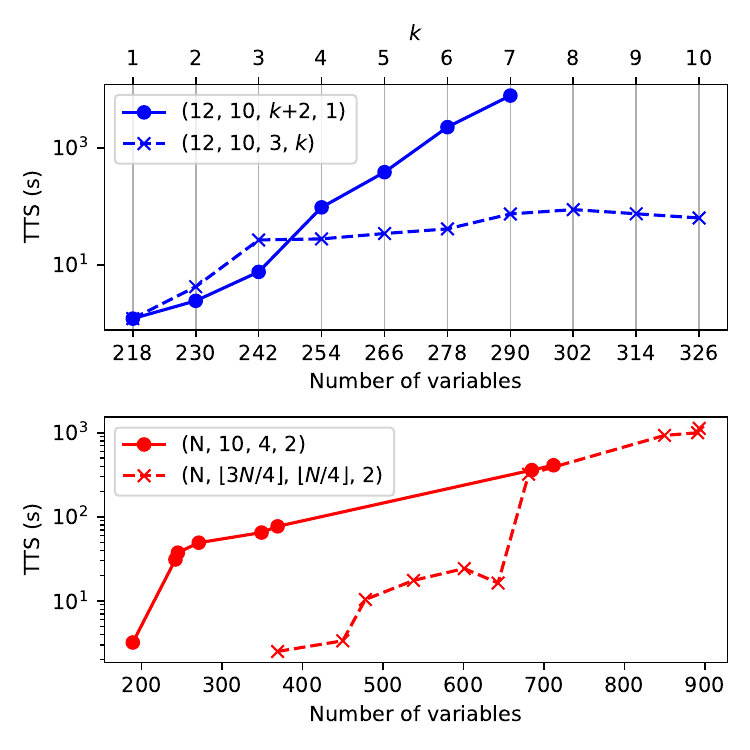}
    \caption{Time required by CPLEX to find the exact solution for several instances. The upper panel shows the scaling $vs$ the number of backbone nodes (dots) and frequencies (crosses). In the lower panel, we show the scaling as a function of the number of nodes.}
    \label{fig:opt_time}
\end{figure}

Without aiming to a complete numerical analysis of our problem, we provide some trends that we observed during our calculations on selected instances. 
We start by reporting the time required for the optimization in Fig. \ref{fig:opt_time}. In the upper panel, we describe how the computational time varies as we increase $B$ and vary $F$. To this aim, we selected two kind of instances characterised by $(12, 10, k+2, 1)$ and $(12, 10, 3, k)$ for $k = 1, ..., 10$, which corresponds to varying $B$ and $F$ respectively, and generated as described previously. Since the underlying graph is always the same, the number of overlaps $N_{\rm 3-over}$ is unchanged and the number of variables scales in the same way with $k$ in both cases. We can immediately see that increasing the number of backbone nodes $B$ leads to an exponentially increasing optimization time. This means that, in order to obtain the exact solution within a practical TTS, $B$ should be as small as possible. A different trend is observed in $F$, for which we observe an initial exponential increase followed by a plateau which basically signals that all the frequency conflicts leading to interference are solved and adding additional frequencies only increases the degeneracy of the solution space. 

In the lower panel, we show how the time needed to find the solution increases as $N$ is increased.
Since $N_{\rm 3-over}$ depends on the graph structure, the number of variables changes non-linearly as $N$ is increased. We consider the two cases $(N, 10, 4, 2)$, which corresponds to distributing 10 antennas over the sites (full red dots), and  $(N, \lfloor 3N/4 \rfloor, \lfloor N/4 \rfloor, 2)$, which corresponds to simultaneously increasing the number of sites and available drones (red crosses). The latter is a more realistic case, since the number of drones has to increase with $N$ in order to cover a significant number of sites. In analogy with the case of increasing $B$, we observe an exponential increase of the optimization time as the number of variables is increased. However, we observe that the optimization time may dramatically vary depending on the considered instances. In particular, we notice jumps in the optimization time, which correspond to increasing value of $B$. When the value of $\lfloor N/4 \rfloor$ reaches $B=4$, the two lines converge to a similar exponential scaling.

\begin{figure}[t]
    \centering
    \includegraphics[width=0.5\textwidth]{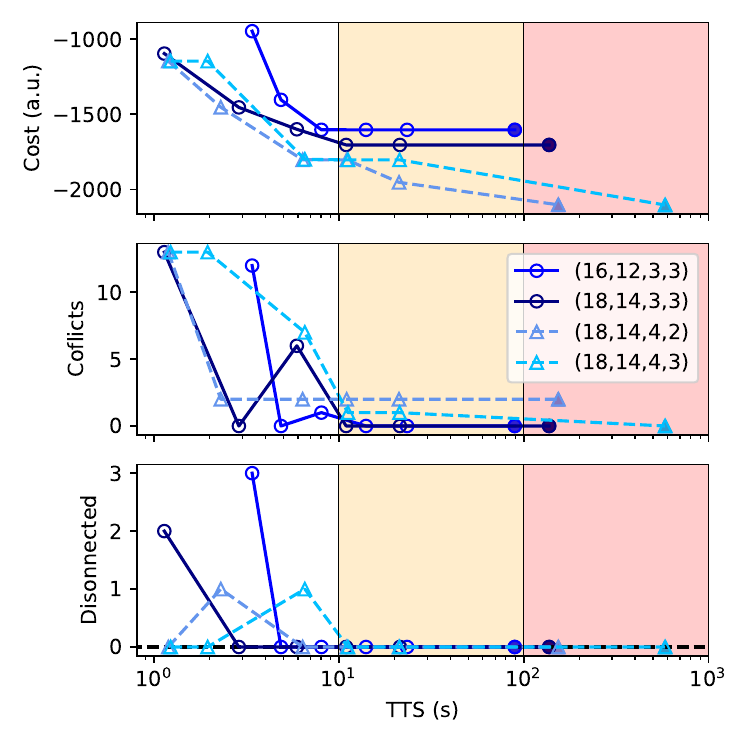}
    \caption{Metrics of the solution for different instances by fixing a maximum optimization time. The horizontal axis shows the Time-to-solution (time to create the instance plus optimization time). The last point for each instance is obtained by letting the solver run until the optimal solution is found (full markers).
    The upper panel shows the total cost of the found (suboptimal or optimal) solution, the central panel the number of interference conflicts, while the lower panel displays the number of disconnected leaves.}
    \label{fig:cost_vs_time}
\end{figure}

Finally, we investigate the performances of CPLEX in cases in which the network has to adapt to changes in the scenario. This situation involves the solution of the optimization problem at each time-step. Reasonable time to refresh the positions of the UAVs are between 10 and 100 seconds, following the movement of the rescuers on the territory. This means that the we have to solve an optimization problem within that time window.
In Fig. \ref{fig:cost_vs_time} we show the convergence of the CPLEX algorithm to the exact solution and the quality of the solution as a function of the allowed time window for different problem sizes. The white area highlights TTS lower than 10 seconds, the orange area indicates the range of TTS that is still useful for practical applications (between 10 and 100 seconds), while the red area shows TTS too high for the typical timescales involved in drone redeployment. We run these calculations for instances that are solvable exactly within $10^3$ seconds, in order to have a reference value. To this aim, we set a maximum time and we look at the solution quality after that time (empty markers). We also allow CPLEX to search for the exact solution without time constraints. The corresponding costs are indicated as full markers. We observe that some instances, namely $(16, 12, 3, 3)$ and $(18, 14, 3, 3)$, reach an exact solution within 10 seconds. However, some additional time is required by the solver to assess that the found solution is actually optimal. All the other instances require more time to reach the exact solution. Looking at the generated solutions at intermediate time-steps, however, it is clear that connectivity requirements are generally already reached after the first few seconds for all the instances considered here. The majority of the time seems to be required to optimize the frequency assignment, although more in depth analysis in required to put this observation on a more quantitative ground. 

\section{Solution using quantum adiabatic algorithm}

\begin{figure}
    \centering
    \includegraphics[width=0.5\textwidth]{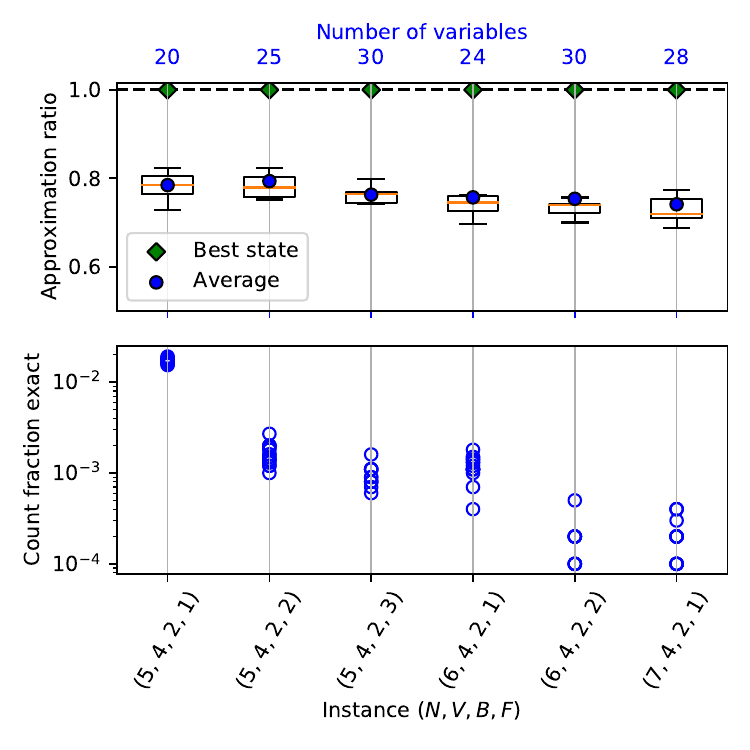}
    \caption{Performance metrics of QAA for a selection of small problem instances that fit into the current exact emulator, which means $N_q \leq 31$. The horizontal axis to the top shows the number of variables $N_q$ while the lower axis the specific instance. The top panel shows the \emph{single-instance} approximation ratio for the best state and the expectation value, with the experiments repeated 16 times. Boxplots show the distribution of the samples for the average. The best measured state coincides with the solution in all the tests so the boxplot is not shown for this quantity. In the lower panel, we show the normalized number of counts of the the solution for the various repetitions. }
    \label{fig:quantum_sol}
\end{figure}

Although deterministic state-of-the-art solvers as CPLEX still outperform their quantum counterparts in solving integer programs of industrial relevance \cite{Tarquini2026, tarquini2026dronedeliverypackingproblem, tarquini2026emergencyhubplacementneutralatom}, the previous analysis shows that they can hardly convey near-to-real-time solutions to certain problems. For this reason, we speculate that at some point quantum computers could be able to return high-quality solutions in the time window required for dynamical re-deployment of the UAVs, in which exact classical solutions are not attainable. For this reason, we use a quantum emulator to solve some small-scale instances of our problem, showing that it is indeed possible to extract high-quality solutions. Specifically, we use the \emph{qiskit} package \cite{Qiskit} in python for the emulation.
In Fig. \ref{fig:quantum_sol}, we show the results obtained using the QAA for some instances. The method adopted here is the same described in Ref. \cite{Vandelli2024}. The number of binary variables that we could handle for this kind of experiment was $N_q=30$, which corresponds to the number of qubits. For this reason, we solve the problem without the cubic term.
We emulate a noiseless finite-measurement experiment with $10^4$ shots and we repeat it 16 times to collect sufficient statistics. 
We investigate two relevant metrics for this algorithm, namely the approximation ratio $\alpha(E) = 1 - \frac{E-H_{\rm min}}{H_{\rm max}-H_{\rm min}} $ and the count fraction of the exact solution. The upper panel of Fig. \ref{fig:quantum_sol} shows the approximation ratio for the average (blue dots) and for the best solution (green diamond). We also added a boxplot to show the finite-measurement variability. Although we did not optimize the hyper-parameters of the algorithm, we are able to recover the exact solution for all the instances considered here. 
In the lower panel of Fig. \ref{fig:quantum_sol}, we show the probability of the exact solution. We observe that this probability is rather low, especially for larger $N$, but does seem to decrease dramatically with the problem parameters, at least for the instances considered here.  

\section{Conclusions}

In this work, we present an original Integer Quadratic Programming formulation of the c-BUP problem. This problem aims to identify \emph{ad-hoc} wireless networks based on 5G technologies using inter-connected UAVs to provide optimal signal service on the ground. Solutions to this problem can help rescuers and first-aid agencies in a natural disaster-stricken territory to efficiently communicate with command and control base stations.

Mathematically, we tackle a complex problem defined on two different, yet related, graphs, incorporating the following components: (1) maximizing ground signal coverage, (2) alternating assignment of neighboring frequencies to avoid local interferences (weighted graph coloring problem formulation), (3) minimizing overlap regions characterized by signal links with the same frequencies, and (4) coupling terms and constraints that enforce air-to-air LoS connectivity of the UAVs network.

We solved the overall problem using the well-established CPLEX solver by IBM. We analyzed the exact solutions generated with CPLEX and the scaling of the optimization time for various selected instances. We observed the exact solution time displaying an exponential scaling with respect to the number of sites $N$ and backbone nodes $B$, while exhibiting a plateau as a function of $F$. 
Besides, we studied the performances of the CPLEX solver in finding suboptimal solutions of the problem instances within a given time window in which we expect the positions of the rescuers at ground not to vary substantially in real scenarios. 

Interestingly, CPLEX finds the exact solution within this time constraint only for the smallest problem instances investigated here. However, a shorter time is often needed to find suboptimal solutions with the desired UAVs connectivity properties, while most of the computing time is spent in optimizing ground coverage and interference. 
As a consequence, our study showcases that our IQP formulation of the c-BUP problem, supplemented with currently available commercial solvers like CPLEX, can support the deployment of UAV wireless networks involving up to 50 drones and up to 10 different frequencies in current emergency response operations. 
However, the advent of beyond-5G technologies and the integration of Internet-of-Things (IoT) devices are anticipated to significantly increase the number of drones utilized in future rescue and relief operations. Consequently, we foresee that specialized algorithms implemented on quantum or quantum-inspired hardware may eventually be capable of finding (sub)optimal solutions much more quickly than currently available classical solvers. To this end, we propose a QUBO formulation of the problem, which is particularly well-suited for implementation on solvers such as simulated annealing and quantum algorithms. These algorithms are compatible with execution on both quantum computers and quantum annealers.
We run some preliminary calculation using the QAA on a quantum emulator and we showcase the ability of the algorithm to recover the exact solution of the problem for a few small instances. Further investigations are needed in order to get a thorough picture of the performances of this and other quantum algorithms on the c-BUP problem.

\section*{Acknowledgments}
The authors express their sincere gratitude to Dr. Gennaro Raiola and Tommaso Alongi who contributed to the discussions during the preparation of this manuscript. 

\section*{Note}

Certain aspects of the work described in this manuscript are the subject of a patent application filed by Leonardo S.p.A.

\bibliographystyle{unsrturl}
\bibliography{example}

\end{document}